\documentclass{emulateapj}

\sfcode`A=1000 \sfcode`B=1000 \sfcode`C=1000 \sfcode`D=1000
\sfcode`E=1000 \sfcode`F=1000 \sfcode`G=1000 \sfcode`H=1000
\sfcode`I=1000 \sfcode`J=1000 \sfcode`K=1000 \sfcode`L=1000
\sfcode`M=1000 \sfcode`N=1000 \sfcode`O=1000 \sfcode`P=1000
\sfcode`Q=1000 \sfcode`R=1000 \sfcode`S=1000 \sfcode`T=1000
\sfcode`U=1000 \sfcode`V=1000 \sfcode`W=1000 \sfcode`X=1000
\sfcode`Y=1000 \sfcode`Z=1000

\slugcomment{{\sc Accepted by ApJ Letters, 2014 October 14}}

\shorttitle{G2 during Periapse Passage}
\shortauthors{Witzel et~al.}

\usepackage{graphicx}
\usepackage{natbib}
\usepackage{amsmath}

\begin{document}

\title{Detection of Galactic Center Source G2 at 3.8 microns during Periapse Passage}
\author{
G. Witzel\altaffilmark{1}, A. M. Ghez\altaffilmark{1}, M. R. Morris\altaffilmark{1}, B. N. Sitarski\altaffilmark{1}, A. Boehle\altaffilmark{1}, S. Naoz\altaffilmark{1}, R. Campbell\altaffilmark{2}, E. E. Becklin\altaffilmark{1}, G. Canalizo\altaffilmark{3}, S. Chappell\altaffilmark{1}, T. Do\altaffilmark{4}, J. R. Lu\altaffilmark{5}, K. Matthews\altaffilmark{6}, L. Meyer\altaffilmark{1}, A. Stockton\altaffilmark{5}, P. Wizinowich\altaffilmark{2}, S. Yelda\altaffilmark{1}
}
\altaffiltext{1}{Dep. of Physics and Astronomy, University of California Los Angeles (UCLA), 465 Portola Plaza,  Los Angeles, CA 90095}
\altaffiltext{2}{W. M. Keck Observatory, 65-1160 Mamalahoa Hwy., Kamuela, HI 96743}
\altaffiltext{3}{Dep. of Physics and Astronomy, University of California Riverside, 900 University Ave., Riverside, CA 92521}
\altaffiltext{4}{Dunlap Institute for Astronomy and Astrophysics, University of Toronto, 50 St. George Street, Toronto, M5S 3H4, ON, Canada}
\altaffiltext{5}{Institute for Astronomy, University of Hawaii, 2680 Woodlawn Drive, Honolulu, HI 96822}
\altaffiltext{6}{Division of Physics, Mathematics, and Astronomy, California Institute of Technology, Pasadena, CA 91125, USA}

\email{ghez@astro.ucla.edu}

\begin{abstract}


We report new observations of the Galactic Center source G2 from the W. M. Keck Observatory. G2 is a dusty red object associated with gas that shows tidal interactions as it nears closest approach with the Galaxy's central black hole.  Our observations, conducted as G2 passed through periapse, were designed to test the proposal that G2 is a 3 earth mass gas cloud. Such a cloud should be tidally disrupted during periapse passage. The data were obtained using the Keck II laser guide star adaptive optics system (LGSAO) and the facility near-infrared camera (NIRC2) through the {\it K'} [2.1 $\mu$m] and {\it L'} [3.8 $\mu$m] broadband filters. Several results emerge from these observations: 1) G2 has survived its closest approach to the black hole as a compact, unresolved source at {\it L'}; 2) G2's {\it L'} brightness measurements are consistent with those over the last decade; 3) G2's motion continues to be consistent with a Keplerian model. These results rule out G2 as a pure gas cloud and imply that G2 has a central star. This star has a luminosity of $\sim$30 $L_{\odot} $ and is surrounded by a large ($\sim$2.6 AU) optically thick dust shell. The differences between the {\it L'} and Br-$\gamma$ observations can be understood with a model in which {\it L'} and Br-$\gamma$ emission arises primarily from internal and external heating, respectively. We suggest that G2 is a binary star merger product and will ultimately appear similar to the B-stars that are tightly clustered around the black hole (the so-called S-star cluster).

\end{abstract}


\section{Introduction}

Gillessen et al. (2012) reported the detection of a very red infrared object approaching the supermassive black hole (SMBH) at the Galactic Center on an orbit with a predicted closest approach of only 3000 times the radius of the event horizon.  Their detection of Brackett-gamma emission led them to interpret it as a dusty, $3$ earth mass ($\rm{M}_{\oplus}$) gas cloud. If this object (G2) is indeed a gas cloud, it would be disrupted by the tidal forces of the SMBH during closest approach and some of it would be accreted (\citealt{2012ApJ...750...58B}; \citealt{2012ApJ...755..155S}; \citealt{2012ApJ...759..132A}). Consequently, G2 has generated tremendous interest since it can be followed through the predicted accretion event and, possibly, provide new insight into accretion physics (\citealt{2013arXiv1311.4507S}; \citealt{2014MNRAS.440.1125A}; \citealt{2013ApJ...768..108S}; \citealt{2012ApJ...752L...1M}; \citealt{2013MNRAS.432..478S}; \citealt{2014ATel.6242....1H}; \citealt{2014ATel.6247....1C}; \citealt{2012Natur.481...51G,2013ApJ...763...78G,2013ApJ...774...44G}).  

The identification of G2 as a pure gas cloud, however, is controversial.
In the pure disrupting gas cloud scenario, G2 is required to have formed relatively recently ($\sim$1995-2000), close to the moment when it first became detectable with new adaptive optics technologies, and at a position that is well inside the formal apoapse position (\citealt{2012ApJ...750...58B}). Since this makes the unusual demand that G2 has been fortuitously observed during the exact decade of its entire existence, many alternative models containing a central stellar source have been proposed (e.g., \citealt{2012NatCo...3E1049M}, \citealt{2012ApJ...756...86M}, \citealt{2012RAA....12..995M}, \citealt{2013ApJ...768..108S}, \citealt{2013ApJ...776...13B}, \citealt{2014A&A...565A..17Z}; see also \citealt{2014ApJ...786L..12G}).  The presence of a central star would allow G2 to survive its closest approach and would not demand a recent formation event.  It would also, in most scenarios, reduce the amount of gas expected to be accreted onto the central black hole following closest approach (e.g., \citealt{2014arXiv1401.0553F}).

Observationally, very little is known about G2. It has been imaged as a very red point source in the near-infrared with adaptive optics systems\footnote{which are necessary to isolate G2 from other sources in this crowded region of the Galaxy}, where it has been detected at wavelengths of 3-5 $\mu$m ({\it L'} - M), but is very faint at 2 $\mu$m ($m_{K'}<20$ mag, \citealt{2013ApJ...773L..13P}, \citealt{2012Natur.481...51G}; see also \citealt{2013A&A...551A..18E}).  Spectroscopically, it is a faint emission-line object, best detected in Br-$\gamma$ (\citealt{2013ApJ...773L..13P}, \citealt{2013ApJ...774...44G}), which shows a slightly elongated, rather compact core and some low surface brightness emission that appears to form leading and trailing tidal tails.  The interpretation of exactly how much of the low surface brightness ionized gas is associated with G2 is complicated by the highly structured gas streams, also seen in emission, that abound in the projected vicinity of Sgr A* (\citealt{2012Natur.481...51G,2013ApJ...763...78G,2013ApJ...774...44G}; \citealt{2013ApJ...773L..13P}, \citealt{2014IAUS..303..264M}).  

With a predicted closest approach having occurred in Spring 2014, the pure gas model can now be tested. In this model the {\it L'}-band component of G2 is interpreted as dust embedded in the gas. It should follow the spatial evolution of the Br-$\gamma$ component, and thus, lose its compactness at {\it L'}. In this paper, we present new {\it L'} imaging observations of G2 at the predicted moment of closest approach and during the following few months.

\section{Observations}
New near-infrared images of G2 were obtained on 2014 March 20,  May 11, July 3, August 4 and 5, 2006 May 21, and 2005 July 30 using the LGSAO system (\citealt{2006PASP..118..297W}, \citealt{2006PASP..118..310V}) and NIRC2 (P.I. K. Matthews) at the W. M. Keck Observatory as part of our long-term study of the central supermassive black hole and its environs (\citealt{1998ApJ...509..678G,2008ApJ...689.1044G}). 
At the time of our observations, G2 and SgrA*, the emissive source associated with the black hole, are expected to be spatially unresolved from each other in our NIR observations. 
The observational set-up used during these measurements enables us to disentangle the emission of G2 from that of Sgr~A*, which is a highly variable source at infrared wavelengths (\citealt{2003Natur.425..934G},  \citealt{2004ApJ...601L.159G}, \citealt{2004A&A...427....1E}, \citealt{2007ApJ...667..900H}) .

Each night, images were obtained in NIRC2's narrow field mode (10 mas/pix), interleaving observations through the {\it K'} [2.1 $\mu$m] and {\it L'} [3.8 $\mu$m] broadband filters. Individual exposure times of 28 sec (10 coadds $\times$ 2.8 sec) and 30 sec (60 coadds $\times$ 0.5 sec) at {\it K'} and {\it L'}, respectively, resulted in a duty cycle time of 134 sec for the two-wavelength cycle.   To optimize the efficiency of operations and performance, the images on G2 were obtained in a stare-mode described in \cite{2007ApJ...667..900H} and the {\it L'} sky images were taken over a range of rotator angles according to \cite{2010ApJ...718..810S}. Table 1 summarizes all of the new data collected, and the historic data used in this analysis.


\section{Analysis}

All the images were analyzed with standard image reduction techniques, as laid out in detail for our group's earlier work (e.g., \citealt{2008ApJ...689.1044G}, \citealt{2009ApJ...690.1463L}, \citealt{2014ApJ...783..131Y}).   Each image was sky-subtracted, flat-fielded, bad-pixel and image-distortion corrected.   
All the individual images from a given night and at the same wavelength were then averaged to obtain a deep combined map for each night, and were de-convolved frame by frame with a Lucy-Richardson algorithm (\citealt{1974AJ.....79..745L}) to obtain time-series of flux density for both bands. To determine statistical uncertainties beyond the formal fitting errors, we additionally created three sub-maps from three simultaneous subsets of frames.


Since G2 and Sgr~A* are spatially unresolved, we disentangled the measurement of their brightness spectrally.  At {\it K'}, the source at G2's predicted location is assumed to be dominated by Sgr~A*\footnote{G2 in all earlier measurements is fainter than Sgr~A* by at least factor of ten ($m_{K'\rm{G2}} <$ 20; \citealt{2013ApJ...773L..13P}, \citealt{2013A&A...551A..18E}, \citealt{2014IAUS..303..264M}).}.  At {\it L'}, the source is expected to be the combination of G2 and Sgr~A*.   While Sgr~A*'s brightness is highly variable, our interleaved {\it K'} measurements of Sgr~A*, and the well measured and constant {\it K'}-{\it L'} color for Sgr~A* (\citealt{2007ApJ...667..900H}; \citealt{2014IAUS..303..274W}) allow Sgr A*'s {\it L'} flux to be estimated and removed. The details of this approach are described below.

Photometric estimates of G2 and Sgr~A* are extracted using two different approaches: point-spread-function (PSF) fitting and aperture photometry on deconvolved images.  In the first method, {\it StarFinder}, a PSF-fitting program (\citealt{2000SPIE.4007..879D}), is used to identify and characterize point sources in each combined map, and in the corresponding sub-maps. This resulted in astrometric and photometric values for Sgr~A* in {\it K'} and for the unresolved G2/Sgr~A* source in {\it L'}.  In the second method, aperture photometry is carried out on the individual deconvolved frames. The PSF for the deconvolution process is obtained by running {\it StarFinder} on individual frames. The restoring beam had the FWHM half the resolution at each wavelength, and the aperture diameter was 60 and 120 mas for {\it K'} and {\it L'}, respectively.  The aperture photometry values are obtained at each wavelength by averaging over the resulting time-series of flux densities.  While this second procedure was intended to assist with any confusion with additional sources that might be near G2 or Sgr~A*, the photometry in both approaches agree well with each other. The final values are the average of the two approaches, and the differences are treated as an additional, albeit negligible, source of uncertainty.

Photometric calibration was accomplished with non-variable calibrators (\citealt{2007ApJ...659.1241R}, \citealt{2010ApJ...718..810S}) in the immediate surrounding of Sgr~A*. The exact set of stars used is identical to that in \cite{2008ApJ...689.1044G} for PSF-fitting results, and to that in \cite{2012ApJS..203...18W} and Witzel et al. (in prep.) for the results from aperture photometry on deconvolved images. The different sets were chosen to guarantee comparability between the results here and in earlier measurements, but both establish consistent zero-points.  

To infer the flux density of Sgr~A* in {\it L'} from our {\it K'} measurements, we applied the known spectral index and the extinction values in the infrared. Sgr~A*'s spectral index has been shown to be constant with brightness and constant in time to within $\Delta \alpha=$0.1 (\citealt{2014IAUS..303..274W}), and we adopt a value of $\alpha = 0.6 \pm 0.2$, which includes the systematics of the extinction correction (\citealt{2007ApJ...667..900H}, \citealt{2014IAUS..303..274W} ) and which is based on the same photometric calibration applied here. For this analysis, an extinction law 
published by \cite{2010A&A...511A..18S}, and zero-points from \cite{2000asqu.book..143T} 
were applied. We subtract the inferred {\it L'}-band, reddening-corrected flux density for Sgr~A* from the reddening-corrected value for the combined G2 + Sgr~A* point source.

The final G2 brightness values are reported without reddening correction, for ease of comparison with earlier observed photometry. We report the photometric values for each individual observation in 2014 as well as the variance-weighted average. The uncertainties for the individual nights incorporate the statistical errors of the zero-point calibration and the sub-map photometry. The average value additionally includes the systematic differences between the PSF-fitting and the deconvolution method, the error of the spectral index, and the uncertainties of the extinction values.

An upper limit on G2's size for each 2014 epoch is obtained from radial profiles of G2's emission. To create the radial profile from the epochal images the inferred {\it L'} flux of Sgr~A* and the closest neighboring stars (S0-2, $m_{L'} \sim 12.7$; S0-8, $m_{L'} \sim 13.9 $) are removed through PSF subtraction. The positional and brightness information for S0-2 and S0-8 are obtained from the {\it StarFinder} analysis discussed above. The exact location of Sgr~A* is best inferred from a dataset in which Sgr~A* is particularly bright. Since Sgr~A* was faint ($<$ 2 mJy at {\it K'} dereddened) during all of the {\it K'}/{\it L'} interleaved datasets, we make use of the {\it K'}-band data obtained on 2014 August 5 when Sgr~A* was bright (7 mJy, dereddened). This allows for a positional accuracy for Sgr~A* of $\sim$ 1 mas, which is dominated by the reference frame alignment uncertainty between the dates. The radial profile of G2 is compared with the PSF and the reduced $\chi^2$-value is calculated. Additionally, to derive an upper limit for the size of G2 the PSF is convolved with a Gaussian. The upper limit is the FWHM of the Gaussian that corresponds to a $\chi^2$ probability of $p = 0.003$. The centroid of G2 with SgrA* and nearby sources removed gives the position of G2, which can be determined to within ~10 mas (uncertainty dominated by dust contributions; see \citealt{2013ApJ...773L..13P}).

\section{Results}

\begin{figure*}
\centering
\includegraphics[scale=0.6, angle=0]{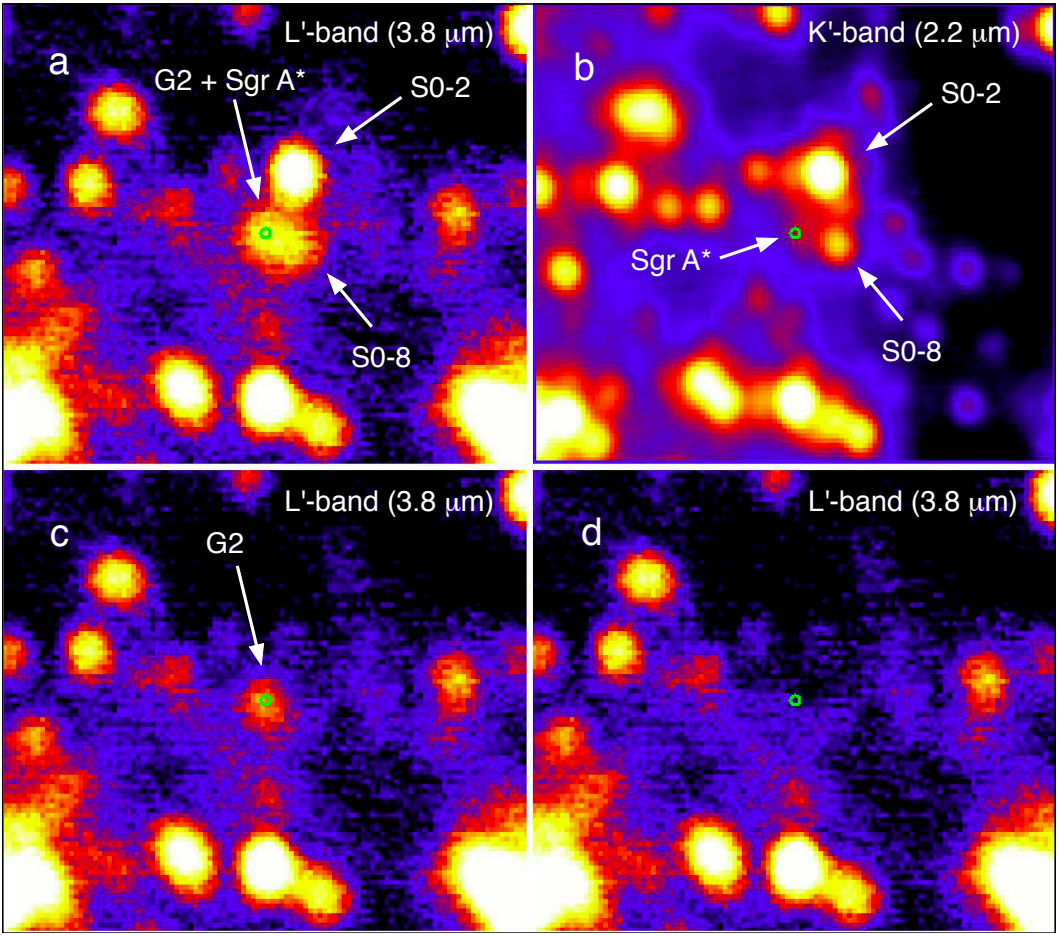}
\caption{ A 1'' x 1'' region of {\it L'} (3.8 $\mu$m; a,c,d) and {\it K'} (2.1 $\mu$m; b) images centered on Sgr~A*.  These images are constructed from data obtained on 2014 March 20.  The {\it L'} image (a) shows the combined flux of Sgr~A* and G2, which are unresolved in these observations. The {\it K'} image (b) shows that Sgr~A* is in a low emission state.  Fig.~c shows the point source subtracted image (with \mbox{S0-2}, S0-8, and Sgr~A* removed) that reveals G2.  Fig.~d shows the same image after subtracting a PSF scaled to the inferred flux density of G2 at the position of G2. The clean result implies compactness of G2's spatial structure. The green circle depicts the position of Sgr~A*.}\label{KLmaos}
\end{figure*}

\begin{figure}
\centering
\includegraphics[scale=0.30, angle=0]{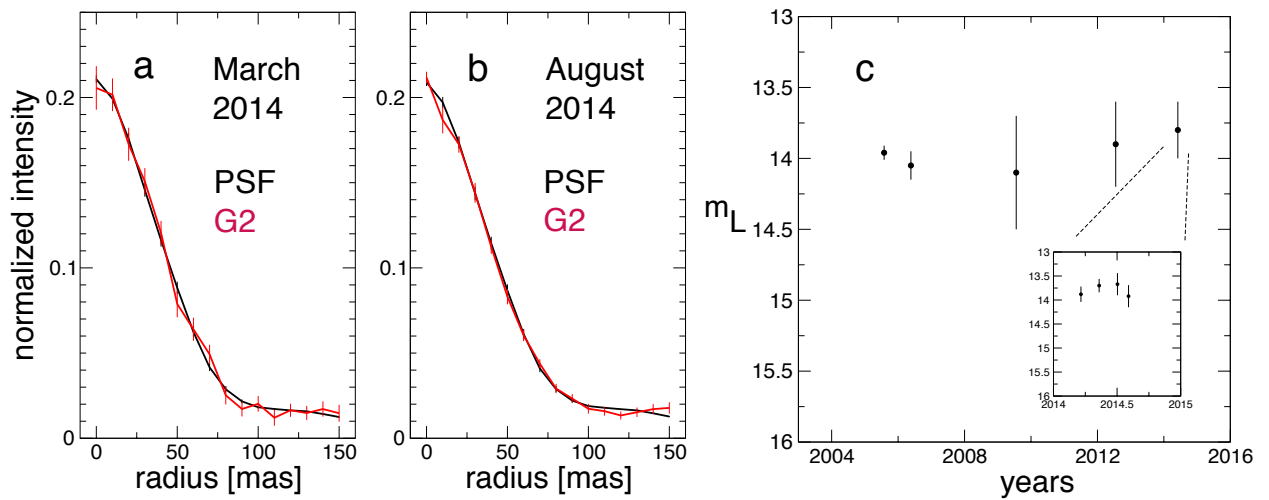}
\caption{ Fig~a,b: Radial plots of G2 and the PSF, shown for the high quality 2014 March and August datasets. In the March epoch Sgr~A* was at its minimum flux of the four datasets presented in this work and, therefore, this radial plot has the smallest systematic uncertainty due to the subtraction process of Sgr~A*. The error bars represent the binning statistics. G2 is fully consistent with a point source. Fig.~c: Photometry of G2 over the last 9 years. The inset depicts the individual measurements in 2014.}\label{s}
\end{figure}

Fig.~\ref{KLmaos} shows that G2 is easily detected in our {\it L'} images. The following key results emerge from the analysis of these images:

\begin{itemize}

\item
{\it G2 has survived its closest approach to the central black hole as a compact, unresolved source at {\it L'}}. Our observations took place very near to G2's closest approach of 2014 March 16 ($\pm$ 2 months) and extended well after this date (5 months). All images reveal G2 and there is no evidence for this object to be extended at {\it L'} in any 2014 epoch, as implied by the $\chi^2$ values in Tab.~\ref{tab2}. We place a $3\sigma$ upper limit on the diameter of G2's emission at {\it L'} of $32$ mas, which correspond to 260 AU at a distance of $\sim 8$ kpc (see Tab.~\ref{tab2} and Fig.~\ref{s}a,b).

\item
{\it G2's L' brightness measurements are consistent with those made over the past decade.}
As the inset to Fig.~\ref{s}b displays, all our G2 {\it L'} brightness estimates from 2014 are in agreement with one another. Furthermore, the 2014 average value of $13.8 \pm 0.2$ is also consistent with earlier measurements (Fig.~\ref{s}b). Tab.~\ref{tab2} summarizes the quantitative results for G2's brightness.

\item
{\it G2's motion continues to be consistent with a Keplerian orbit model.} The {\it L'} position of G2 in 2014 is consistent with the predictions of our orbital model (\citealt{2013arXiv1312.1715M}). Within these predictions, the new position favors orbital solutions with shorter periods ($< 500$ yr) and reduces the uncertainty in the lower range of the periapse distance $a_{\rm{min}}$ by a factor of 3 ($a_{\rm{min}}$ = 215 $\pm$ 30 AU).
\end{itemize}

\section{Discussion and conclusions}

Our {\it L'} measurements present a very different view of G2 from what has been seen through the Br-$\gamma$ line emission measurements.  The Br-$\gamma$ line emission traces hot gas ($\rm{T}_{e^-} \sim 10^4$K; \citealt{2012Natur.481...51G}) that appears to be {\it externally} heated by ionizing photons from massive stars in the vicinity of G2. In Br-$\gamma$, G2 shows clear evidence of tidal interaction with the black hole (\citealt{2012Natur.481...51G,2013ApJ...763...78G,2013ApJ...774...44G}, \citealt{2013ApJ...773L..13P}, \citealt{2014arXiv1407.4354P}).  This interaction and its evolution have been seen both in the increasing linewidth associated with the spatially compact `head' as well as the increasing extent of the low surface brightness tail(s) associated with G2. However, it is important to note that the Br-$\gamma$ line emission measurements imply only that {\it some} gas associated with G2 has a size that exceeds its tidal radius. 


Unlike Br-$\gamma$, the {\it L'} emission remains spatially unresolved, continues to follow a well-defined Keplerian orbit, and is constant in brightness.  This allows us to rule out a pure gas cloud model as such a gas cloud model predicts that G2's brightness and size should undergo substantial changes (e.g., \citealt{2012ApJ...759..132A}, \citealt{2014arXiv1407.4354P}). Instead, G2's {\it L'} emission appears to be coming from an optically thick dust shell surrounding an underlying star. Two lines of evidence point to this conclusion.  

\begin{figure}
\centering
\includegraphics[scale=0.4, angle=0]{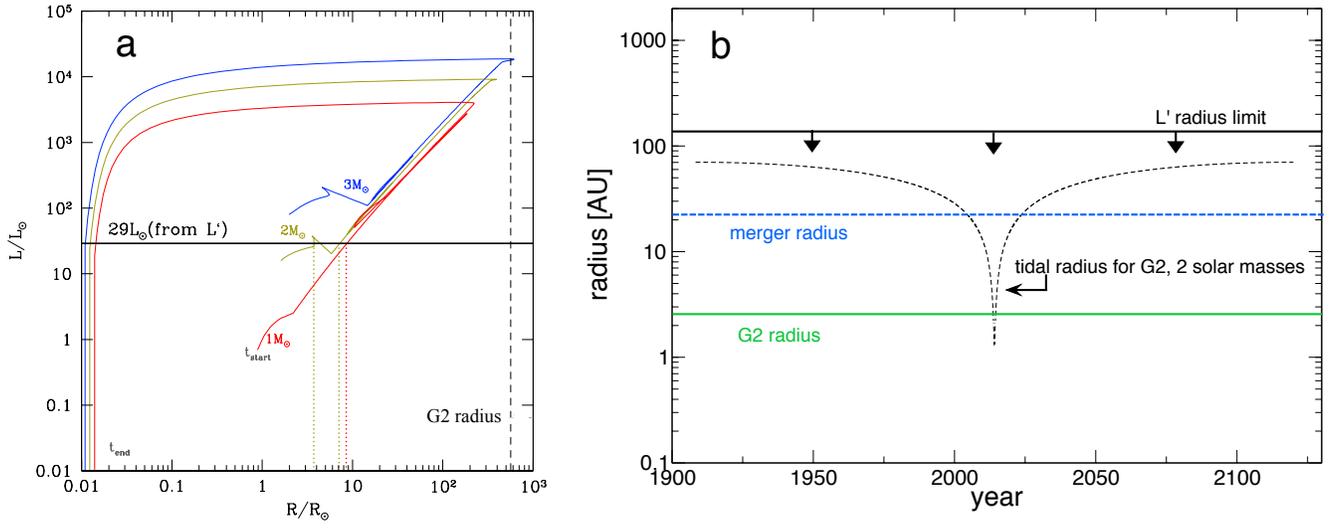}
\caption{Fig.~a: Stellar luminosity as a function of radius. We consider three
different masses, $1,2$ and $3$~M$_\odot$ and evolve them according to
the stellar evolution code SSE (\citealt{Hurley+00}). Over-plotted is the
luminosity of G2 as measured from {\it L'} emission. Fig.~b: Tidal radius of G2 as a function of time assuming as mass of $\sim$2 solar masses, and the radius of G2 derived for a optically thick blackbody. For comparison we show the {\it L'} radius limit derived directly from our observations and the radius of a model binary star merger.}\label{model}
\end{figure}

First, the following facts are now established:
\begin{itemize}
\item The {\it L'} flux has been invariant at $\sim 2.1$ mJy (reddening-corrected, see Tab.~\ref{tab2}) since 2005, in spite of a factor of 10 change in distance from the central black hole.
\item In 2004, G2 had an {\it L' - M'} color of $\sim 0.3$ (reddening-corrected, \citealt{2012Natur.481...51G}), which corresponds to a blackbody temperature of $\sim$ 560~K.
\item No {\it K'} detection has been made in any epoch.
\item The {\it L'} emission is much more compact than the emission in Br-$\gamma$, and thus originates in a different region.
\end{itemize}
We conclude that this is most likely explained by optically thick dust that has had constant temperature and size over the past decade and is {\it internally} heated.  Modeling the dust emission as a blackbody results in an inferred luminosity $L_{G2} = 29 \; L_{\odot}$, which can easily be generated by a 2~$M_{\odot}$ main-sequence star or a somewhat lower mass post- or pre-main-sequence star  that is temporally very close to the main-sequence (see Fig.~\ref{model}a); higher mass stars are ruled out with the inferred luminosity. 

The challenge is to explain the unusually large size inferred from the blackbody. This size is $\sim$2.6~AU and is $\sim 100$ times larger than the photospheric size of 30~$L_{\odot}$ stars. Several scenarios can account for such a large size, but can be excluded based on other properties. A single young star surrounded by a protoplanetary disk, as proposed by \cite{2012NatCo...3E1049M}, would have to be observed edge-on, which is rather unlikely. The corresponding face-on systems would not appear as such highly obscured objects. A common envelope surrounding a binary system containing a giant star would be too luminous.  

As we had suggested in \cite{2013ApJ...773L..13P} and was followed up by \cite{2014arXiv1405.6029P}, a binary merger is a natural model for G2. As Fig.~\ref{model}b shows, G2 has a radius in the range calculated for several putative stellar merger products (\citealt{2013A&A...555A..16T}; also \citealt{2010A&A...522A..75K} and \citealt{2013ApJ...777...23Z}). A second line of evidence that G2 indeed is a $\sim$ 2~$M_{\odot}$ merger product comes from tidal radius arguments. The constancy of the {\it L'} flux and the compactness of G2 lead us to conclude that the tidal radius of the source, which is proportional to its distance from the black hole, has not become smaller than the source size for most of the time. For a black hole mass of $M_{\rm{BH}} = 4.3 \cdot 10^{6} M_{\odot}$, the tidal radius is
\begin{equation}\label{tidalr}
r_t = 1.31 \; \rm{AU} \cdot \frac{R_{\rm{3D}}}{215 \; \rm{AU}}  \times \left(\frac{M_{\rm{G2}}}{2M_{\odot}}\right)^{1/3}  \; \; ,
\end{equation}
where $R_{\rm{3D}}$ is the 3D-distance of G2 from the black hole. Fig.~\ref{model}b shows the time development of the tidal radius $r_t$. For a mass of $\sim$2~$M_{\odot}$ the derived {\it L'} size of G2 does not show any tidal interaction with the black hole, except possibly near periapse passage. This is consistent with G2's {\it L'} size and photometry not evolving during its approach to the black hole. We note that the mass of G2, in the case in which it is indeed a merger product, can be different from the main-sequence star mass assumed here. However, the tidal radius is only weakly dependent on the mass (a factor 10 in mass corresponds to a factor $\sim 2.2$ in tidal radius).

In this picture, in which G2's {\it L'} emission originates within the tidal radius, gas and dust beyond the tidal radius are removed by the tidal forces, creating extended optically thin tidal tails. While this extended emission is faint at {\it L'},  it is competitive with the central source at Br-$\gamma$.  Thus, the Br-$\gamma$ emission is a by-product of the tidal interaction whereas the {\it L'} emission traces the properties of the merger product. 

The interpretation of G2 as a merger product suggests that it will eventually look like a typical member of the S-star cluster after the extended atmosphere contracts on a Kelvin-Helmholtz time scale. The environs of the central supermassive black hole might be particular conducive to mergers as close binary stars in orbit around the SMBH could experience Kozai oscillations that increase the eccentricity of their orbits (\citealt{2014arXiv1405.6029P} and references therein; \citealt{Naoz+13}). If this mechanism is indeed at work, it could dramatically increase the rate of stellar mergers. The plausibility of such binary interactions
contributing to the central stellar population relies on the poorly known distribution and fraction of stars in binary
systems at the GC (e.g., \citealt{2014arXiv1405.6029P}). Further investigations of the stellar dynamics are underway to address the question whether it is possible that an important fraction of the S-stars has resulted from such mergers.


\acknowledgments
We wish to dedicate this paper to Gerry Neugebauer (1932-2014).

\vspace{0.5cm}

We thank the anonymous referee. Support for this work was provided by NSF grant AST-0909218 and AST-1412615, the Levine- Leichtman Family Foundation, the Preston family graduate fellowship (held by AB), and the Janet Merrot Student Travel Awards.  The W. M. Keck Observatory is operated as a scientific partnership among the California Institute of Technology, the University of California and the National Aeronautics and Space Administration. The Observatory was made possible by the generous financial support of the W. M. Keck Foundation.

\clearpage

\begin{table*}
\begin{center}
\caption{Observations of G2 and Sgr~A*}\label{tab2}
\begin{tabular}{llllllllll}
\tableline
\tableline
&&&&&&&&&\\
& & \multicolumn{2}{c}{Data Quality} &  \multicolumn{3}{c}{reddening corrected flux density} & obs. Photometry & \multicolumn{2}{c}{radial profile} \\
&&&&&&&&&\\
Date & $\rm{N}_{frames}$ & $\langle\rm{FWHM}\rangle$ & $\langle\rm{Strehl}\rangle$ & Sgr~A* & Sgr~A* & G2 + Sgr~A*  & G2 & $\chi^2$/DOF & size limit\\
& {\it K'}/{\it L'} & {\it K'}/{\it L'} & {\it K'}/{\it L'} & $S_{K'}$ [mJy] & $S_{L'}$ [mJy] &  $S_{L'}$ [mJy] & $m_{\rm{{\it L'}}}$ && AU \\
&&&& measured & inferred & measured & inferred &&\\
\tableline
&&&&&&&&&\\
\multicolumn{10}{c}{Primary Data} \\
\tableline
&&&&&&&&&\\
2014 March 20 & 22/21 & 67/91 & 0.21/0.44 & 1.1 $\pm$ 0.1 & 1.5 $\pm$ 0.2 & 3.7 $\pm$ 0.4 & 13.88 $\pm$ 0.16&0.48 & $<$370\\
2014 May 11 & 9/9 & 67/90 & 0.21/0.44 & 1.5 $\pm$  0.2 & 2.2 $\pm$ 0.2 & 4.7 $\pm$ 0.4 & 13.70 $\pm$ 0.14& 1.08 & $<$350\\
2014 July 3 & 64/64\footnote{(58/58) in the case of aperture photometry on the individual deconvolved frames.} & 70/108 & 0.19/0.32 & 1.6 $\pm$ 0.3 & 2.3 $\pm$ 0.4 &  4.9 $\pm$ 0.5 & 13.67 $\pm$ 0.23&1.14 & $<$380\\
2014 August 4 & 28/28 & 63/92 & 0.20/0.42 & 1.8 $\pm$ 0.2 & 2.5 $\pm$ 0.4  & 4.6 $\pm$ 0.6 & 13.92 $\pm$ 0.23&0.65 & $<$260\\
&&&&&&&&&\\
average &&&&&&& 13.8 $\pm$ 0.2 &&\\
&&&&&&&&&\\
\tableline
&&&&&&&&&\\
\multicolumn{10}{c}{Auxiliary Data} \\
\tableline
&&&&&&&&&\\
2005 July 30 & -/56 & -/81 & -/0.66 &&&& 13.96 $\pm$  0.05&&\\
2006 May 21 & -/19 & -/82 & -/0.56 &&&& 14.05  $\pm$ 0.10&&\\
2009 July 22 & 4/- & -/86 & -/0.46 &&&& 14.1 $\pm$  0.4&&\\
2012 July 20 - 23 & 1314/- & -/91 & -/0.48 &&&& 13.9  $\pm$ 0.3&&\\
&&&&&&&&&\\
2014 August 5 &  127/- &  57/-  & 0.26/- & 7.0 $\pm 0.3$ &&&&& \\
&&&&&&&&&\\
\tableline
\end{tabular}
\label{data}
\end{center}
\end{table*}

\end{document}